\documentclass[ps,prb,twocolumn,showpacs]{revtex4}

\usepackage{epsfig} \usepackage{color}

\begin{document}
\bibliographystyle{prsty}  \title{Efficient readout of flux qubits at
degeneracy} \author{A.\ K\"ack} \affiliation{Dept. of Microtechnology
and Nanoscience  (MC2), Chalmers University of Technology, 41296
Gothenburg, Sweden}

\author{F.K.\ Wilhelm} \email{wilhelm@theorie.physik.uni-muenchen.de}

\affiliation{Department Physik, ASC, and CeNS, Ludwig-Maximilians-University,
Theresienstr.\ 37, 80333 M\"unchen, Germany}

\begin{abstract}
We present a superconducting circuit consisting of a flux qubit and a
single-charge transistor serving as a detector. As flux and charge are
conjugate, the transistor can detect states of the qubit close to the
flux degeneracy point, when the eigenstates are quantum superpositions
of fluxes. The coupling has a flip-flop symmetry conserving the
total number of excitations, and so 
the measurement outcome results in the absence or presence of an
incoherent tunneling cycle. We evaluate the performance of a practical
device and show that it is an attractive tool for measuring at the 
degeneracy point.
\end{abstract}


\maketitle

Small Josephson junction circuits are a prototype system for studying
generic quantum effects in solid-state physics \cite{Makhlin01}. 
These
devices are a candidate for 
scalable quantum computing, which requires full and precise control of 
quantum states in real time. The strong development
in this field has lead to a number of substantial achievements such as
superpositions \cite{vanderWal00} and real-time control
\cite{Chiorescu03} of collective states as well as  the control of
charge states \cite{Nakamura99} and the demonstration of a
controlled-not gate \cite{Yamamoto03}. A highly attractive realization of a superconducting
qubit is based on magnetic flux states \cite{Mooij99}. 

A crucial and subtle issue in this effort is the proper understanding and
design of the read-out device. Quantum measurements are intrinsically
invasive \cite{Braginsky95}. 
Ideally,
the back-action exclusively consists of 
instantaneous projection on
eigenstates of the measured observable. In general, and specifically for 
quantum systems such as the Josephson circuits, 
the back-action consists of two parts: The dephasing process
which projects onto the measurement basis, 
and relaxation which induces transitions between these
states  leading to errors in the observed variable. The former is
inevitable and is an integral part of quantum control protocols,
whereas the latter has to be  avoided \cite{Makhlin01,PRBR03C}. This aspect
makes quantum nondemolition (QND) schemes highly attractive. Thus,
for fundamental as well as for practical reasons we are faced with a
problem of understanding and ultimately engineering the properties of
the read-out device on the quantum level. 

Next to the backaction of the detector, any quantum bit realized in a
solid-state environment  suffers from decoherence induced by the other degrees of
freedom of the solid-state environment and of the fluctuations of the control fields --- flux noise in
the case of flux qubits. 
Slow fluctuations from the ubiquitous $1/f$ noise in
the solid-state environment are particularly destructive, here resulting mostly
in critical current fluctuations. 
Thus, it is very attractive to operate devices at the charge degeneracy point where
they are to first order immune to these  fluctuations \cite{Vion02,Chiorescu03,Majer02}. It is possible
to achieve a very stable intrinsic bias at that point by flux trapping
\cite{Majer02}.
Fixing the bias to that point can lead to restrictions on the control parameters. Although operation by AC-pulses is still possible, it is
not evident how to read out the state at the end. In a flux-based
device operated at flux degeneracy any
Hamiltonian-dominated\cite{Makhlin01,PRBR03C} measurement cannot
distinguish the energy eigenstates of the system. The natural solution
is the measurement of a different variable such as phase
\cite{Vion02}, which in the particular case of Ref.\ [\onlinecite{Vion02,Chiorescu03}] 
still requires to move
out of the optimum point.

In this letter, we are going to show how, using a Cooper-pair transistor, the state of the qubit can
be read out with high fidelity and low unwanted 
backaction without leaving the
degeneracy point. The physical coupling mechanism is based on 
geometric phases.
By appropriate biasing of the transistor and
by using a readout channel which filters out unwanted processes, 
the physical processes are flip-flop transitions between qubit and
detector, which lead to a QND-like and quantum limited backaction. 
We will discuss values of  experimentally accessible parameters.

At the degeneracy point, the energy eigenstates are coherent
superpositions of flux states, which cannot be discriminated by flux 
detection. On the other hand, the  conjugate variable, the
charge, diminishes its intrinsic quantum fluctuations and can in principle be
detected.

\begin{figure}
\includegraphics[width=0.9\columnwidth]{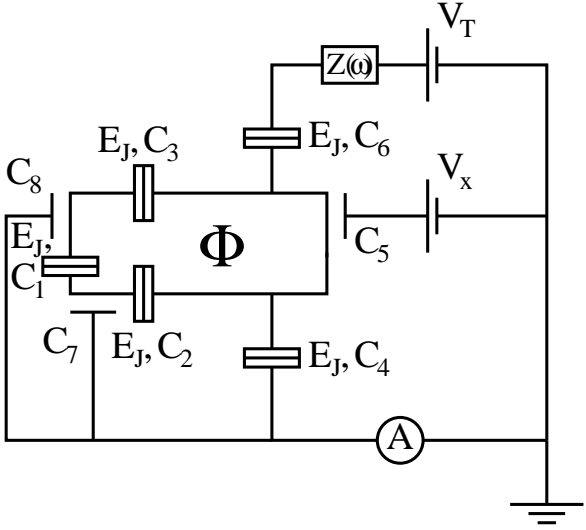}
\caption{Proposed circuit for the read out of a flux qubit (three
junction loop on the left, junctions 1,2 and 3, controlled by the external flux $\Phi$) at the degeneracy point using an SET (part
on  the right, other junctions). In order to establish the interaction, the qubit has
to have (stray) capacitances to ground, $C_4$, $C_7$, and $C_8$ as well.
\label{fig:circuit}}
\end{figure}

The circuit we envisage is depicted in Fig.~(\ref{fig:circuit}).  It
consists of a three-junction loop frustrated by a magnetic flux close
to half a flux quantum, following Ref.\ [\onlinecite{Mooij99}],  one of
the islands of which also serves as the island of a single-charge
transistor. I.e., it is coupled to voltage sources by small
Josephson junctions exhibiting Coulomb blockade, and is subject to an
electrostatic gate voltage.  For fixing the reference, the
different parts of the device are connected to ground either
capacitively or galvanically.

In the limit we are interested in, both components are in their
generic regimes, i.e.\ $E_{\rm J,qubit}\gg E_{\rm c, qubit}$ 
$E_{\rm c, SET}\gg E_{\rm J, SET}$. Here and henceforth, $E_J$ and $E_c=2e^2/C$ are the
Josephson and charging energies of the respective junction.
The interaction between flux and charge is via their
mutual  geometric phases.  Charge moving around magnetic flux
acquires an Aharonov-Bohm phase, a flux line moving around a charge
acquires an Aharonov-Casher phase  \cite{Elion93,Friedman02}.
This analogy is a manifestation of vortex-charge duality.

In order to accommodate loop constraints and Kirchhoff rules, we start from 
a Lagrangian analysis. The Lagrangian
of such a coupled circuit can be schematically represented  in terms
of the junction phases $\phi_i$ as  ${\cal L}\left(\lbrace
\phi_i\rbrace, \lbrace \dot{\phi}_i\rbrace \right)={\cal L}_{\rm
ch}-{\cal L}_{\rm J} - {\cal L}_{\rm S}$. The charging part reads
${\cal L}=\frac12\left(\frac{\Phi_0}{2\pi}\right)^2 \dot{\vec{\phi}}^T
{\bf C}\dot{\vec{\phi}}$ where ${\bf C}$ is the capacitance matrix and
the phase velocities have been arranged in a column vector.  The
Josephson part reads ${\cal L}_J=-\sum_j E_{J,i}\cos\phi_i =
U_J(\phi_1,\phi_2,\phi_3)$ where  the Josephson energies are given
through the critical currents by $E_{J,i}=\Phi_0I_{c,i}/2\pi$. The
work done by the voltage sources, finally,  is counted as ${\cal
L}_S=-\frac{\Phi_0}{2\pi}\vec{V}^T{\bf C_g}\dot{\vec{\phi}}$.  In our
case the capacitance matrix is, using the notation from
Fig.~(\ref{fig:circuit})
\begin{equation}
  \label{eq:capacitances}
  C= \left(
  \begin{array}{ccc}
    C_8+C_1+C_3 & -C_1 & C_8\\ -C_1 & C_1+C_2+C_7 & C_7 \\ C_8 & C_7 &
    C_7+C_8 + C_{\Sigma}
  \end{array} \right),
\end{equation}
where $C_\Sigma=C_4+C_5+C_6$ is the total capacitance of the
transistor, and the source term is given by
\begin{equation}
  \label{eq:sources}
  \overrightarrow{Q_g}= C_g \overrightarrow{V} = \left(
  \begin{array}{c}
    0\\ 0\\ -C_5 V_x -C_6 V_T
  \end{array} \right)
\end{equation}
From here, we can find the canonical variables and the Hamiltonian
\begin{equation}
  \label{eq:Hamiltonian}
  {\cal H} = \frac{1}{2}
  \left(\frac{2\pi}{\Phi_0}\right)^2(\overrightarrow{P}-\frac{\Phi_0}{2\pi}\overrightarrow{Q_g})^T
  C^{-1}(\overrightarrow{P}-\frac{\Phi_0}{2\pi}\overrightarrow{Q_g})
  -U_J(\phi_1,\phi_2,\phi_3).
\end{equation}
The canonical momentum $\overrightarrow{P}$ is given by
$\overrightarrow{P} = \left(\frac{\Phi_0}{2\pi}\right)^2C
  \dot{\overrightarrow{\Phi}} + \frac{\Phi_0}{2\pi}
  \overrightarrow{Q_g}$. We want to rewrite this Hamiltonian now as
$H=H_q+H_{\rm q, SET}+H_{SET}$.
From Eq.~(\ref{eq:Hamiltonian}) we see that the coupling between the
transistor and the qubit has the form of $-i\partial_{\phi_k}
P_3$. This coupling term can, as well as the coupling to voltage sources, 
be transformed into phase factors 
by a gauge transformation $\psi(\vec{\phi})\rightarrow\psi (\vec{\phi})\exp\left(-i\int \hat{P_3}d\vec{\phi}\right)$. This
results in a qubit Hamiltonian in phase representation of the same shape as
Ref. \onlinecite{Orlando99}, with a slightly modified capacitance
matrix. Performing the two level approximation leads to $H_q^\prime=\frac12\left(\matrix{\epsilon&\Delta \cr \Delta^\ast
  &-\epsilon\cr}\right)$ with $\Delta=|\Delta| e^{i\chi}$ where
$\hat{\chi}=\delta\phi \hat{P}_3$ and $\delta\phi$, the separation of the minima in the potential
landscape corresponding to the persistent current states\cite{Mooij99}. In
this gauge the charging part of the transistor island has its
standard form
$ E_C(n_L-n_R-n_g)^2 + 2eV( \kappa n_L + (1-\kappa)n_R)$
, where $n_{L/R}$ is the number of charges that have tunneled through the left/right junction and $n_L-n_R$ is the number of extra charges on the transistor island, $\kappa = C_R/(C_L+C_R)$ will be assumed to be $1/2$ through the rest of this paper. $C_L=C_6$ and $C_R=C_4+C_5+C_6+C_7+C_8$ are the effective left/right capacitances of the transistor and $n_x$ is the dimensionless gate voltage. 
As $E_{\rm c,SET}\gg E_{\rm J,SET}$,
we can perform a two level approximation also for the transistor island. In order to simplify the treatment of the external
Josephson junctions, we observe that the phase $\chi$ measures the charge tunneling through the SETs
and obtain the 
qubit-transistor interaction:
\begin{equation}
(H_q+H_{\rm q,SET})^\prime=\frac12\left(\matrix{\epsilon&\Delta e^{i\phi N}\cr \Delta
e^{-i\phi N}&-\epsilon\cr}\right).
\end{equation}
Here $\phi$ is a constant set by the geometric capacitances of the circuit.
At the flux degeneracy point, we have $\epsilon=0$. 
By performing the unitary transformation
$\hat{U}=\exp\left(\frac{i\phi}{2}\hat{\sigma}_z\otimes\hat{\tau}_z\right)$,
where $\sigma_z$ is a Pauli matrix operating in in the qubit subspace and
$\tau_z$ is a Pauli matrix operating in in the transistor subspace, we
finally obtain
\begin{widetext}
\begin{eqnarray}
\label{eq:fullH}
\hat{H}&=&\frac{\Delta}{2}\hat{\sigma}_z +\frac{\delta E_{\rm
ch}}{2}\hat{\tau}_z
+E_{J,SET}\hat{1}\otimes\cos\phi\left[\hat{\tau}_+(\hat{l}+\hat{r})+\hat{\tau}_-(\hat{l}^\dagger+\hat{r}_i^\dagger)\right]
+iE_{J,SET}\hat{\sigma}_x\otimes\sin\phi\left[\hat{\tau}_+(\hat{l}+\hat{r})-\hat{\tau}_-(\hat{l}^\dagger+\hat{r}_i^\dagger)\right]\nonumber\\
&& +\frac12 \left(V_T/2+\sum_i
\lambda_i(\alpha_i+\alpha_i^\dagger)\right)\hat{n}_l+\frac{1}{2}
\left(-V_T/2+\sum_i\mu_i(\beta_i+\beta_i^\dagger)\right)\hat{n}_r+\sum_i
\omega_i\alpha^\dagger_i\alpha_i+\sum_i\omega_i\beta^\dagger_i\beta_i.
\label{eq:heff}
\end{eqnarray}
\end{widetext}

Here, we already have assumed to operate at the flux degeneracy point
$\epsilon=0$ and we have changed to the energy basis such that
$\hat{\sigma}_z$ now describes energy and $\hat{\sigma}_x$ flux
eigenstates. $n_{L/R}$ count charges having tunneled to the left and
right lead respectively and $l/r$ transfer charges from the left/right lead to
the island. Moreover, $\hat{\tau}_{\rm\pm}$ are the raising and
lowering operators for the island charge. As we envision the readout 
to happen completely on the supercurrent branch (see below), we omit quasiparticle terms. 
For finding dissipative action and finally observing a voltage signal,
one has to include resistors described by the 
$\alpha$
and $\beta$ operators, which are modeled as baths of harmonic
oscillators along the lines of the familiar $P(E)$ analysis of 
the Caldeira-Leggett model \cite{PRL01}.

This is the starting point for our rate equation study of the readout
process. The gauge transformation has turned Aharonov-Casher
into Aharonov-Bohm phases. 
If the qubit is in one of the flux, $\hat{\sigma}_x$, eigenstates, the
environment will not induce any transitions of the qubit and the qubit
state will only influence the phases of the two tunneling events. Such 
phase differences cannot be detected in the present geometry, as this
would require a closed superconducting loop. 
Notably, qubit and island charge formally appear to be
uncoupled, but are jointly affected by  coupling to the
superconducting leads. This is a result of our gauge transformation:
In the previous gauge, the qubit eigenstates had different phases 
depending on the eigenstate of $\hat{\tau}_z$, thus  if the island charge 
described by $\hat{\tau}_z$ flips
the qubit can also make a transition. Note, that the qubit energy 
splitting has the same value in both $\hat{\tau}$ states. 

\begin{figure}
  \centering \includegraphics[width=0.45\textwidth]{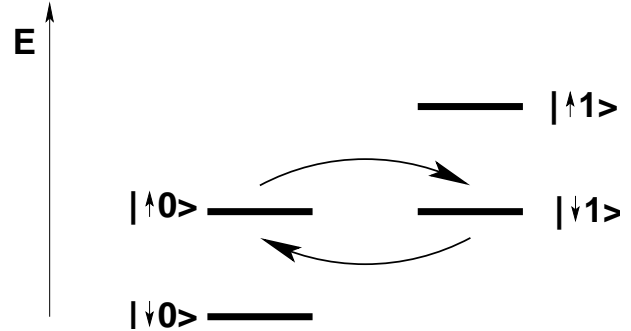}
  \caption{The eigenlevels of the unperturbed Hamiltonian. The arrows
  indicate the wanted transitions. The up/down arrows indicate the
  qubit states and the numbers 1/0 indicate the transistor states.}
  \label{fig:levels}
\end{figure}

When $E_{J, SET} < E_{c,SET}$ and the
dimensionless conductance of the junction in the transistor is small,
we can treat the tunneling term in the Hamiltonian in
Eq.~(\ref{eq:fullH}) perturbatively. 
The first two terms of Eq.\ (\ref{eq:fullH}) plus the free leads in the
last two terms constitute the unperturbed part. 
The device has to be biased at 
$\Delta = \delta E_c$. The
eigenlevels of the unperturbed system are shown in
Fig.~(\ref{fig:levels}).
In the ideal case, the parameter $\phi$ in Eq.~(\ref{eq:fullH}) is
exactly $\pi/2$. In this case tunneling through the junctions of the
transistor is only possible if, at the same time, the qubit also
switches. As a charge tunnels through first the left, then the right
junction (corresponding to following the right-going arrow in
Fig.~(\ref{fig:levels}), then the left-going) a charge has been
transfered through the transistor, while the qubit and the transistor
returns to the original state.  This lead to a quantum flip-flop
process where the flips of the transistor island as current passes are accompanied
by opposite transitions --- flops --- of the qubit: Charging the island relaxes the qubit,
discharging the island excites the qubit. All other transitions are suppressed due to
symmetry. Thus we get a current running through the transistor if the
qubit is in the excited state, but no current in the ground state. By
reading out the current running through the transistor the state of
the qubit can be detected. As there is no mixing, we can take as much
time to read out the state as we want. Remarkably, if after the measurement
the transistor island is uncharged, the post-measurement state is the qubit eigenstate
corresponding to the measured observable without relaxation. This is the same outcome as 
in a QND measurement, even though the Hamiltonian does {\em not} satisfy the
QND condition \cite{Braginsky95} and the qubit {\em does} make 
transitions during the operation. This result is owed to the flip-flop
symmetry of the Hamiltonian. 
The tunneling probability per
second is constant leading to a Poissonian current distribution, which
means that arbitrary large resolution is available in this
ideal limit by reading out during arbitrarily long time.

An important note is that if the qubit is initially in a superposition of its
eigenstates (remember, that eigenstates at the degeneracy point are themselves equal superpositions
of current states), a single
charge cycle in the transistor will destroy the coherence of the qubit.
As a single charge cycle can in principle be detected with the method of
Ref.\ \onlinecite{Astafiev04}, this is a quantum-limited measurement. 

For the non-ideal case, the parameter $\phi$ in eq.\ (\ref{eq:fullH}) is not exactly $\pi/2$
and we have unwanted transitions that lead to mixing of the qubit. As
$\phi$ depends on the capacitor setup, adjusting it with high precision
appears to be
difficult. In that case, there is a risk of making a transition to the
$|\downarrow 0\rangle$ state by spontaneous emission to the environment, 
see Fig.\ \ref{fig:levels},
which halts the detection process and potentially falsifies the result. 
Thus, we need to find a detection scheme which suppresses spontaneous emission,
i.e.\ which has an effective spectral density peaked around zero. 

This can be done by using incoherent Cooper pair tunneling\cite{Berkeley}\cite{IngoldNazarov}.
By putting an impedance in series with the transistors the
supercurrent of the transistor is suppressed. The fluctuations of the
voltage across the junction leads instead to an incoherent transfer of
Cooper pairs. Still, the energy of the Cooper pair condensate defines
a preferred tunneling window and leads to the desired band-pass property.
The strong coupling this setup provides and the large qubit splitting
compared to the temperature of the oscillator bath makes the bandpass
properties quite accentuated which combined with the symmetry
properties should result in a good signal to noise ratio.

\begin{figure}[h]
  \centering
  \includegraphics[width=0.7\columnwidth]{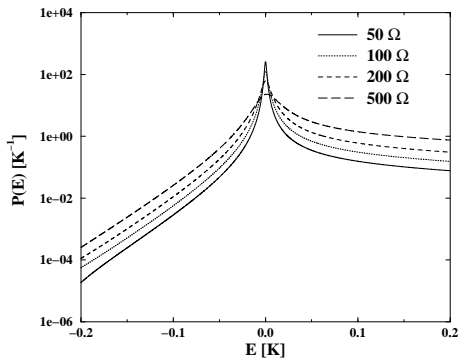}
  \caption{P(E) for an ohmic environment for different resistance of the environment. The junction capacitance used was 0.25 fF and the temperature was 25 mK. Note the lin-log scale.}
  \label{fig:p_e}
\end{figure}

The incoherent rates can be calculated using standard $P(E)$ theory
for an Ohmic environment \cite{PRL01}, using 
\begin{equation}
P(E)=\cases{0 & $E < -k_BT/g$\cr
1/T& $|E|<k_BT/g$\cr
2/gE & $E\gg k_BT/g$}
\end{equation}
where $g$ is the dimensionless conductance of the environment. An example of P(E) is shown in Fig. \ref{fig:p_e}.
Thus, the rates for the wanted processes are $\Gamma\simeq \sin^2(\phi)/T$ 
whereas the unwanted one is $\Gamma_{\rm error}\simeq   \cos^2(\phi) 2/gE$
where $E\simeq 2E_c$ is the energy gain of a transition
to the dark $|0\downarrow\rangle$ state. This
way, we can transfer a number of $N=\Gamma/\Gamma_{\rm error}=(gE/2T) \tan^2(\phi)$
Cooper pairs through the device before an error occurs, setting
an error window for the measurement time of $t_{\rm error}\simeq N/I$ where $I$ is the average
current.

Using the results from Fig. \ref{fig:p_e} with a resistance of 50 $\Omega$, this ratio can be estimated to several thousand Cooper pairs. Considering present day technology where single Cooper pairs can be counted \cite{Astafiev04}, this can be straightforwardly detected.



This setup has several advantages
compared to other setups. For instance it allows for a flux qubit read
out on the flux degeneracy point as well a read out for arbitrary
small qubit inductance. Because it uses geometric phases to couple the
qubit and the pointer it does not require the qubit to move into a
mixed charge-flux regime\cite{Amin}. This should make the qubit insensitive
to charge noise while at the same time maintaining  insensitivity to
flux noise to first order.  The physics behind the coupling is also
quite general meaning that it should be possible to adapt this read
out scheme to any system containing conjugate variables.
%
%
%

In this letter, we have demonstrated a
novel and promising read out scheme for flux qubits using the
conjugate variable. It allows for read out on the flux
degeneracy point, leading to improved noise insensitivity. Using
incoherent Cooper pair tunneling as a form of bandpass filtering the
two different qubit states can be distinguished and detected. The
strong coupling provided by the setup allows the bandpass filter to,
in the low temperature regime, provide a good signal to noise ratio. 
We have identified an ideal working point in which the measurement has
QND-like properties and is quantum-limited.

This work was sponsored by the EU through the 
MANAS Marie Cure training site and the IST-SQUBIT2 project. FKW acknowledges support of the DFG 
through SFB 631 and ARDA and ARO under contract number P-43385-PH-QC. 


\end{document}